\documentclass[final,5p]{elsarticle}

\usepackage{lineno,hyperref}
\usepackage{verbatim}
\usepackage{amsmath}
\usepackage{graphicx}
\usepackage{subfigure}
\usepackage{multicol} 

\usepackage{multirow}
\usepackage{array}
\usepackage{subfigure}
\usepackage{booktabs}
\usepackage{amsfonts}
\usepackage{epstopdf}
\usepackage{float}
\usepackage{color}

\modulolinenumbers[5]

\journal{Neural Networks}

\begin{document}

\begin{frontmatter}

\title{Probabilistic Inference of Binary Markov Random Fields in Spiking Neural Networks through Mean-field Approximation}

\author[mymainaddress,mythirdaryaddress]{Yajing Zheng}
\author[mymainaddress,mythirdaryaddress]{Shanshan Jia}
\author[mymainaddress,mythirdaryaddress]{Zhaofei Yu\corref{cor1}}
\author[mymainaddress,mythirdaryaddress]{Tiejun Huang}
\author[mysecondaryaddress]{Jian~K.~Liu}
\author[mymainaddress,mythirdaryaddress]{Yonghong Tian\corref{cor1}}
\ead{yuzf12@pku.edu.cn (Zhaofei Yu); yhtian@pku.edu.cn}

\cortext[cor1]{Corresponding author. }
\address[mymainaddress]{National Engineering Laboratory for Video Technology, Department of Computer Science, Peking University, Beijing 100871, China}
\address[mythirdaryaddress]{Peng Cheng Laboratory, Shenzhen 518055, China}
\address[mysecondaryaddress]{Centre for Systems Neuroscience, Department of Neuroscience, Psychology and Behaviour, University of Leicester, Leicester LE1 7HA, UK}

\begin{abstract}
Recent studies have suggested that the cognitive process of the human brain is realized as probabilistic inference and can be further modeled by probabilistic graphical models like Markov random fields. Nevertheless, it remains unclear how probabilistic inference can be implemented by a network of spiking neurons in the brain.
Previous studies have tried to relate the inference equation of binary Markov random fields to the dynamic equation of spiking neural networks through belief propagation algorithm and reparameterization, but they are valid only for Markov random fields with limited network structure. In this paper, we propose a spiking neural network model that can implement inference of arbitrary binary Markov random fields. Specifically, we design a spiking recurrent neural network and prove that its neuronal dynamics are mathematically equivalent to the inference process of Markov random fields by adopting mean-field theory. Furthermore, our mean-field approach unifies previous works. Theoretical analysis and experimental results, together with the application to image denoising, demonstrate that our proposed spiking neural network can get comparable results to that of mean-field inference.
\end{abstract}


\end{frontmatter}


\section{Introduction}
The human brain is able to process information in the presence of sensory uncertainties~\cite{meyniel2015confidence}. For example, one can easily localize a bird in a tree via noisy visual and auditory cues. Such processes can be understood as probabilistic inference and further modeled by probabilistic graphical models~\cite{koller2009probabilistic, wainwright2008graphical}, including Bayesian networks and Markov Random Fields (MRFs).  With an increasing volume of behavioral and physiological evidence~\cite{knill1996perception,doya2007bayesian,ma2014neural,pouget2016confidence} that humans do actually use probabilistic rules in perception~\cite{kersten2004object,shi2013bayesian}, sensorimotor control~\cite{bays2007computational, kording2004bayesian} and cognition~\cite{chater2006probabilistic,yuille2006vision,jampani2015informed}, probabilistic brain is getting recognized by neuroscientists~\cite{pouget2013probabilistic}. Nevertheless, it remains unclear how the brain can perform inference. Or more precisely, how a network of spiking neurons in the brain can implement inference of probabilistic graphical models?  This problem is of great importance to both computer science and brain science~\cite{Yu2020}. If we known the neural algorithms of probabilistic inference, it is possible to build a machine that can perform probabilistic inference like the human brain. 

In recent studies, many researchers have been devoted to developing neural circuits that can represent and implement inference of undirected probabilistic graphical models, namely MRFs~\cite{koller2009probabilistic}, which is widely used in computational neuroscience~\cite{fischl2002whole,ming2010modeling, probst2015probabilistic,vasta2016hippocampal}.
The reason for focusing on MRFs is that, for directed probabilistic graphical models, one can easily convert them to  MRFs via moralization~\cite{jordan1999introduction,koller2009probabilistic}. 

Here we briefly review these previous studies. 
Litvak and Ullman~\cite{litvak2009cortical} designed neural circuits to implement the operations of summation and multiplication respectively, and further implemented probabilistic computation and inference of MRFs. 
Steimer et al.~\cite{steimer2009belief} proposed using a population of spiking neurons to collect messages and another population to send messages, and then implemented the Belief Propagation (BP) algorithm, a commonly used inference method in probabilistic graphical models~\cite{koller2009probabilistic,wainwright2008graphical}.
All these studies require that each neuron and synapse conduct complicated computation. However, one often observes one basic principle of the neuronal system in the brain that a single neuron or a group of neurons should work in a relatively simple style, while complex functions could be achieved when they are wired together, i.e., collaborated in a network~\cite{Liu2009a, buonomano2009state}.
 
In order to propose biologically more plausible neural networks to implement inference, Ott and Stoop~\cite{ott2007neurodynamics} established a relationship between the inference equation of binary MRFs and the dynamic equation of spiking neural networks through BP algorithm and reparameterization. However, their model relied on the specifically initialized messages and certain topological structures of MRFs. 
Yu et al.~\cite{yu2017neural} went a further step to relax the constraints on initialized messages, but still required the special topological structure and potential function of MRFs. Another important way is based on tree-based reparameterization algorithm~\cite{raju2016inference}, which, however, is only limited to the case of exponential family distributions. 

In this paper, we use a mean-field approximation to treat the inference process of MRFs as a time-continuous system of a recurrent spiking neural network. We analytically prove a precise equivalence between the inference equation of Markov random fields and the dynamic equation of spiking recurrent neural networks. We show that the firing rates of neurons in the network can encode the difference between the probabilities of two states. In addition, we prove that the time course of neural firing rate can implement marginal inference of arbitrary binary Markov random fields. In this way, we can obtain the state of the neuron by counting spikes from each neuron within a time window. We further show that our mean-field approximation unifies the previous approach based on BP algorithm and reparameterization. Theoretical analysis and experimental results, together with an application to the image denoising problem, show that our proposed spiking neural network can get comparable results to that of mean-field inference.

To summarize, our contributions include the following aspects:
\begin{itemize}
\item We propose a spiking neural network model that can implement inference of arbitrary binary Markov random fields. 
\item We prove that there exists a precise equivalence between the dynamics of recurrent neural network and the inference equation of a Markov random field.
\item We show that the previous approach based on BP algorithm and reparameterizations equals mean-field approximation.
\item We show that our proposed spiking neural network can be used to solve practical computer vision problems, like image denoising.
\end{itemize}

The rest of the paper is organized as follows. In section 2 we briefly review MRFs and marginal inference, then we derive the inference equation of MRFs based on mean-field approximation and show how it is related to the dynamic equation of spiking neural networks in section 3. We show the simulation results in section 4 and conclude in section 5. 

\section{Markov Random Fields and Marginal Inference}
In this section, we briefly review MRFs and marginal inference. 
MRFs is one typical undirected probabilistic graphical model that is widely used in computational neuroscience. Thanks to their ability to model soft contextual constraints between random variables, MRFs provide a principled probabilistic framework to model various vision problems~\cite{felzenszwalb2006efficient, chen2007spatio, dong2015simultaneous} since the visual scene modeling usually involves interactions between a subset of pixels and scene components.

In a MRF, a joint distribution $P(\{x\})=P(x_1, x_2, \dots, x_n)$ is defined on the graph, which can be factorized into a product of potential functions according to the structure of the graph. For the MRF in Fig.~\ref{fig:01}, $P(\{x\})$ has the form:

\begin{figure}[thbp]
	\begin{center}
		\includegraphics[width=0.8\columnwidth]{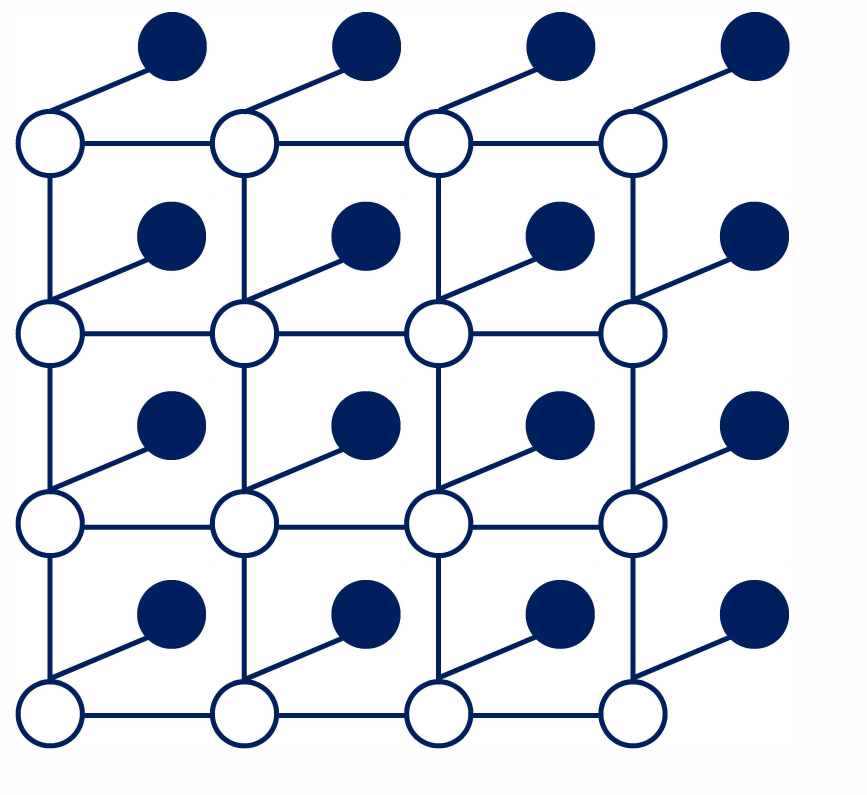}
		\caption{A square lattice pairwise Markov random field. The filled-in circles represent the observed nodes $y_i$, while the empty circles represent the ``hidden" nodes $x_i$.}
		\label{fig:01}
	\end{center}
\end{figure}
\begin{equation}
P(\{x\}) = \frac{1}{Z} \prod_{(i, j) \in E} \Psi_{ij}(x_i, x_j)\prod_{i\in V}\Psi_{i}(x_i, y_i),
\label{eq1}
\end{equation}
where $E$ and $V$ represent the set of edges and nodes in the graph respectively, $\Psi_{ij}(x_i, x_j)$ and $\Psi_{i}(x_i, y_i)$ denote the pairwise and unary potential functions. $Z$ is the partition function defined as 
$$Z = \sum_{x_1, x_2, \dots, x_n}\prod_{(i,j)\in E} \Psi_{ij}(x_i, x_j) \prod_{i\in V}\Psi_i(x_i, y_i).$$ If one defines $J_{ij}(x_i, x_j)=\ln \Psi_{ij}(x_i, x_j)$ and $h_i(x_i)=\ln \Psi_i(x_i, y_i)$
\footnote{As the observed variable $y_i$ is fixed,
one can subsume it into the definition of $h_i(x_i)$.}, Eq.~(\ref{eq1}) can be rewritten as:

\begin{equation}
\label{newjoint}
P(\{x\}) = \frac{1}{Z} \exp \left (\sum_{(i, j) \in E} J_{ij}(x_i, x_j)+\sum_{i\in V} h_i(x_i) \right ) 
\end{equation}

Similar to the studies in \cite{ott2007neurodynamics,yu2017neural}, we assume that $J_{ij}(x_i, x_j) = J_{ij}x_ix_j$ and $h_i(x_i)= h_ix_i$, in which $J_{ij}$ and $h_i$ are constants. 

The inference problems of MRFs include Maximum a Posterior (MAP) estimation and marginal inference. By MAP estimation, we refer to the estimation of a maximum of posterior point estimator. Conversely, marginal inference refers to inferring the posterior or conditional distribution over the latent causes of observations. In this paper, we only consider marginal inference. Specifically, we compute the marginal distribution of each variable $x_i$, 
that is:
\begin{equation}
p_i(x_i) = \sum_{{\bf x} \backslash x_i}P(x_1, x_2, \dots, x_n).
\label{eq3}
\end{equation}

\section{Neural Implementation of Marginal Inference on Binary MRF}
In this section, we will prove that there exists a precise equivalence between the neuronal dynamics of recurrent neural networks and mean-field inference of binary MRFs. We first derive a differential equation that
has the same fixed point as the mean-field inference equation of MRFs, then we show that this differential equation can be easily implemented by the dynamic equation of recurrent neural networks. In the end, we demonstrate that the previous work based on BP algorithm and reparameterization equals the mean-field approximation.

\subsection{Converting Mean-Field Inference into a Differential Equation}
Similar to the studies in \cite{ott2007neurodynamics,yu2017neural}, we only consider inference of binary MRFs in this paper, which means the value of the variable $x_i$ can be 1 or -1 ($x_i=1$ or $-1$). 

As exact inference of MRF is a NP-complete problem \cite{koller2009probabilistic}, approximate inference algorithms like variational methods are often used. The main principle of variational methods is converting the inference problem to an optimization problem:
\begin{equation}
\min _{q({ x})} KL (q(x)||p(x)).
\label{eq4}
\end{equation}
Here the target distribution $p(x)$ is approximated by a simpler distribution $q(x)$, which belongs to a family of tractable distribution. $KL(\cdot)$ represents the Kullback-Leibler divergence between two distributions. 
In the mean-field method, $q(x)$ is set to be a fully factorized distribution, that is $q(x)=\prod_i b_i(x_i)$. By constraining $\sum_{x_i}b_i(x_i)=1$ and differentiating $KL \left(q(x)||p(x) \right)$ with respect to $b_i(x_i)$, one can obtain the mean-field inference equation:
\begin{equation}
b_i^{t+1}(x_i) = \alpha \Psi_{i}(x_i,y_i)\exp \left(\sum_{j\in N(i)}\sum_{x_j}b_j^{t}(x_j)\ln \Psi_{ij}(x_i, x_j) \right),
\label{eq7}
\end{equation}
where $\alpha$ is a normalization constant  to make $\sum_{x_i}b_i(x_i)=1$ and $N(i)$ denotes the set of all neighboring nodes of node $i$. $t$ denotes the number of iterations, and $b_i^t(x_i)$ represents the information received by node $i$ in the $t$ th iteration, which is a function with respect to the state of variable $x_i$. When all the message converge to the fixed point, the marginal probability $p(x_i)$ can be approximated by the steady-state $b_i^{\infty}(x_i)$.
According to the definition $\ln \Psi_{ij}(x_i, x_j)=J_{ij}(x_i, x_j) = J_{ij}x_ix_j$ and $\ln \Psi_{i}(x_i, y_i)=h_i(x_i)= h_ix_i$, Eq.~(\ref{eq7}) can be rewritten as:
\begin{equation}
b_i^{t+1}(x_i) = \alpha \exp \left(\sum_{j\in N(i)}\sum_{x_j}b_j^{t}(x_j)\cdot J_{ij}x_ix_j+ h_ix_i \right).
\label{eq9}
\end{equation}

In order to convert Eq. (\ref{eq9}) to a differential equation, we reparameterize the message $b_i^t(x_i)$ of variable $x_i$ according to:
\begin{equation}
n_i^{t} = b_i^{t}(x_i = 1) - b_i^{t}(x_i = -1),
\label{eq10}
\end{equation}
where $n_i^{t} $ can be seen as the new message received by node $i$ in the $t$ th iteration. Note that here the message $n_i^{t}$ is independent of the state of variable $x_i$. When $n_i^{t} $ converges to the fixed point, it can approximate the probability $p(x_i=1)-p(x_i=-1)$.
Combining Eq.~(\ref{eq9})-(\ref{eq10}) and the condition $b^{t}(x_i=1)+b^{t}(x_i=-1)=1$ defined  on binary MRF, one can get that:
\begin{equation}
\begin{aligned}
n_i^{t+1}  &=b_i^{t+1}(x_i = 1) - b_i^{t+1}(x_i = -1)\\
&=  \alpha \exp \left(\sum_{j \in N(i)}J_{ij} \cdot (b_j^{t}(x_j=1)-b_j^{t}(x_j=-1))+h_i \right) \\
&-\alpha \exp \left(\sum_{j \in N(i)} -J_{ij} \cdot (b_j^{t}(x_j=1)-b_j^{t}(x_j=-1))-h_i \right) \\
&=  \tanh \left(\sum_{j \in N(i)}J_{ij} \cdot (b_j^{t}(x_j=1) - b_j^{t}(x_j=-1)) + h_i \right) \\
& = \tanh \left(\sum_{j \in N(i)}J_{ij} \cdot n_j^{t} + h_i \right).
\end{aligned}
\label{eq16}
\end{equation}
Note that the third equality of Eq.~(\ref{eq16}) holds as

\begin{equation}
\begin{aligned}
 \frac{1}{\alpha} & =\sum_{x_i} \exp \left(\sum_{j\in N(i)}\sum_{x_j}b_j^{t}(x_j)\cdot J_{ij}x_ix_j+ h_ix_i \right) \\
   & = \exp \left(\sum_{j \in N(i)}J_{ij} \cdot (b_j^{t}(x_j=1)-b_j^{t}(x_j=-1))+h_i \right) \\
  &  +\exp \left(\sum_{j \in N(i)} -J_{ij} \cdot (b_j^{t}(x_j=1)-b_j^{t}(x_j=-1))-h_i \right).
\end{aligned}
\end{equation}

It is easy to prove that the following differential equation has the same fixed point as Eq. (\ref{eq16}). 
\begin{equation}
\tau_0 \frac{dn_i(t)}{dt} = -n_i(t) + \tanh \left(\sum_{j\in N(i)}J_{ij} \cdot n_j(t) + h_i \right),
\label{eq12}
\end{equation}
where $\tau_0$ is a time constant that determines the time needed for the network to reach the fixed point. 

\subsection{Dynamic Equation of Spiking Recurrent Neural Networks}
Recurrent neural networks are composed of a population of interconnected neurons, which have been widely used to model cortical response properties in computational neuroscience \cite{rao2004bayesian,rao2005hierarchical}.
Here, we drive the firing-rate based equation of spiking recurrent neural network based on two steps \cite{dayan2001theoretical}: 1) Determining how the total synaptic input to a neuron depends on the firing rate of its presynaptic afferents. 2) Modeling how the firing rate of the postsynaptic neuron depends on its total synaptic input. 

First of all, considering the recurrent neural network consists of $N$ spiking neurons $z_1,z_2,...,z_N$, the input current to the neuron $z_i$ at time $t$ is $I_i(t)$, which includes 
the recurrent input of spike sequence from other neurons and can be computed as:
\begin{equation}
    I_i(t)=\sum_{j=1}^N w_{ij}\int_{-\infty}^t\kappa(t-\tau)S_j(\tau)d\tau,
    \label{eq2}
\end{equation}
where $S_j(\tau)$ denotes the firing spike sequence of neuron $z_j$ defined as a sum of Dirac $\delta$ function $S_j(t) = \sum_f\delta(t-t_j^f)$, $t_j^f$ is firing time of the $f$ th spike of neuron $z_j$. $w_{ij}$ denotes the synaptic weight between neuron $z_i$ and $z_j$, $\kappa(t)$ is the synaptic kernel that describes the time course of the synaptic current in response to a presynaptic spike arriving at time $t$. The most frequently used form of synaptic kernel is an exponential kernel, that is, $\kappa(t)=\frac{1}{\tau_s}\exp \left(-\frac{t}{\tau_s} \right)$ with the membrane time constant $\tau_s$.

In fact, the neural response function $S_j(t)$ could be replaced by the firing rate  $r_j(t)$ of neuron $z_j$ as $r_j(t)=\frac{1}{\Delta t}\int_t^{t+\Delta t}\langle S_j(\tau) \rangle d\tau$ with $\langle S_j(t) \rangle$ denoting the trial-average neural response function, thus Eq. (\ref{eq2}) can be rewritten as:
\begin{equation}
    I_i(t)=\sum_{j=1}^N w_{ij}\int_{-\infty}^t\frac{1}{\tau_s}\exp \left(-\frac{t-\tau}{\tau_s} \right)r_j(\tau)d\tau.
    \label{eq3}
\end{equation}
By taking the derivative of $I_k(t)$ with respect to time $t$, one obtain:

\begin{equation}
\tau_s\frac{dI_i(t)}{dt}=-I_i(t)+\sum_{j=1}^N w_{ij}r_j(t),
    \label{eq4}
\end{equation}
with $\tau_s$ denoting the time constant that describes the decay of the synaptic conductance. 

So far we can determine the input current to postsynaptic neuron
in terms of the firing rates of the presynaptic neurons. To obtain the firing-rate model, we also need to determine the postsynaptic firing rate with the current $I_i(t)$. For time-independent inputs, the firing rate $r_i(t)$ of the postsynaptic neuron $z_i$ can be expressed as $r_i(t)=F(I_i(t))$, where $F(x)$ denotes the neuronal activation function. As the firing rate does not follow changes of the total synaptic
current instantaneously, the firing rate is often modelled by a low-pass filtered version of the synaptic current:

\begin{equation}
    \tau_r\frac{dr_i(t)}{dt}=-r_i(t)+F(I_i(t)).
    \label{eq5}
\end{equation}
Under the constraints of time-independent inputs, the steady state of the postsynaptic current $I_i(t)$ is $\lim_{t\rightarrow \infty}I_i(t)=\sum_{j=1}^N w_{ij} r_j(t)$.  If $\tau_r\gg\tau_s$, we can make the approximation that Eq. \eqref{eq4} comes to equilibrium quickly compared to Eq. \eqref{eq5}.  Consequently, we can further replace $I_i(t)$ by $\sum_{j=1}^N w_{ij} r_j(t)$ in Eq. \eqref{eq5} and obtain:

\begin{equation}
    \tau_r\frac{dr_i(t)}{dt}=-r_i(t)+F \left( \sum_{j=1}^N w_{ij} r_j(t) \right ).
    \label{eq6}
\end{equation}

In recurrent neural networks, except for the input current from recurrent neurons, there also exists an external input current. Incorporating the external input current $I_i^{ext}(t)$ to Eq. \eqref{eq6}, the firing rate of the recurrent neuron $k$ is determined by:

\begin{equation}
    \tau_r\frac{dr_i(t)}{dt}=-r_i(t)+F \left(I_i^{ext}(t)+\sum_{j=1}^N w_{ij} r_j(t) \right).
    \label{eq66}
\end{equation}

\subsection{Implementation of Inference with Neural Network}

Now one can relate the inference equation of MRFs (Eq. (\ref{eq12})) to the dynamics of recurrent neural networks (Eq. (\ref{eq66})).
Obviously, Eq. \eqref{eq12} is equivalent to Eq. \eqref{eq66} if the following equations hold:
\begin{align}
\tau_r &=\tau_0, \label{uh1} \\
r_i(t) &= n_i(t),\label{uh2} \\
w_{ij} &=\left\{\begin{matrix}
~~~J_{ij}~~~~~\text{if}~~j \in N(i)\\ 0~~~~~\text{others}
\end{matrix}\right. , \label{uh} \\
I_i^{ext}(t) &= h_i,\label{uh4} \\
F(x) &= \tanh(x). \label{uh5}
\end{align}

Eq. (\ref{uh1})--(\ref{uh5}) mean that if the synaptic weights $w_{ij}$ and input current $I_i^{ext}(t)$ of a recurrent neural network encode the potential functions $J_{ij}$ and $h_i$ of a binary MRF respectively, the firing rate $r_i(t)$ of neuron $z_i$ encodes the probability  $p(x_i=1)-p(x_i=-1)$.
Moreover, the time course of neural firing rate in the recurrent neural network can implement marginal inference of the MRF. Thus, we can read out the inference result by counting spikes from each neuron within a time window. Note that as the value of $n_i(t)$ varies from $-1$ to $1$, the firing rate $r_i(t)$ in Eq. (\ref{uh2}) could be negative, which is
biological implausible. As discussed in \cite{rao2004bayesian}, we can assume that the actual firing rate $\hat{r}_i(t)$ is linearly related to the "firing rate" $r_i(t)$ obtained from Eq.~(\ref{eq66}), that is, $\hat{r}_i(t)=a r_i(t)+b$. Here $a$ is a positive factor and $b$ is a rectification value that ensure  $\hat{r}_i(t)$ to be positive. In conclusion, we implement mean-field  inference of binary MRFs with spiking recurrent neural networks.

\subsection{Relating Mean-field Inference to Belief Propagation}
\label{relation}
Here we will build the relationship between mean-field inference and BP, and show that the previous work based on BP and reparameterization equals the mean-field inference.

Previous studies have tried to relate BP algorithm of binary MRF to the dynamics of Hopfield Networks by deriving a new formulation of belief propagation based on reparameterization \cite{ott2007neurodynamics,yu2017neural}:
\begin{align}
\label{BP1}
\mu_i^{t+1}&=\tanh \left(  \sum_{j \in N(i)} \tanh^{-1}  \left[ \tanh(J_{ij}) \tanh \left( \sum_{s \in N(j))\setminus i} n_{s \rightarrow j }^t+h_j  \right)  \right] +h_i
\right),
\end{align}
where $\mu_i^{t+1}$ represents the new message after reparameteration of node $i$ at the $t+1$ th iteration, and $\mu_i^{\infty}=p(x_i=1)-p(x_i=-1)$.
$n_{s \rightarrow j }^t$ is a function of the message $m_{s \rightarrow j }^t(x_j)$ in BP that is sent from node $s$ to node $j$ in the $t$ th iteration. To be specific, $n_{s \rightarrow j }^t= \tanh^{-1} \left(m_{s \rightarrow j }^t(x_j=1)-m_{s \rightarrow j }^t(x_j=-1) \right)$. With the assumptions that the number of neighboring nodes of each node is large enough ($N(j)>>1$) and the potential function is small ($J_{ij}<<1$ and $h_i<<1$), Ott et al.~\cite{ott2007neurodynamics} and Yu et al.~\cite{yu2017neural} proved that Eq. (\ref{BP1}) can be simplified to:
\begin{equation}
\label{BP2}
\mu_i^{t+1}  = \tanh(\sum_{j \in N(i)} \tanh(J_{ij}) \cdot \mu_j^{t} + h_i).
\end{equation}
As $J_{ij}<<1$, $\tanh(J_{ij}) \approx  J_{ij}$. Thus one can further simplify Eq. (\ref{BP2}) to:
\begin{equation}
\label{BP3}
\mu_i^{t+1}  = \tanh(\sum_{j \in N(i)} J_{ij} \cdot \mu_j^{t} + h_i).
\end{equation}
One can find that there exists a precise equivalence between Eq.~(\ref{BP3}) and Eq.~(\ref{eq16}), which implies that the previous work based on BP and reparameteration equals the mean-field approximation.
These results suggest that the Hopfield networks used in the previous work actually implement mean-field inference, instead of the BP algorithm. In addition, our current results explain the experiments in~\cite{yu2017neural} where the inference result based on Hopfield networks is not as accurate as that of BP when the potential function is large ($J_{ij}>1$ and $h_i>1$). These errors come from the difference between mean-field inference and the BP algorithm.

\section{Simulation Experiments}
To validate the proposed computational framework, we evaluate the performance of recurrent neural networks through simulation experiments. We firstly test the accuracy of the propose method, and then prove that it is robust to different parameters. At last, we scale up the proposed spiking neural network to solve practical computer vision problems.

\subsection{Testing on the Accuracy of Our Method}
\begin{figure}[t!htb]
	\centering
	\includegraphics[width=0.9\columnwidth]{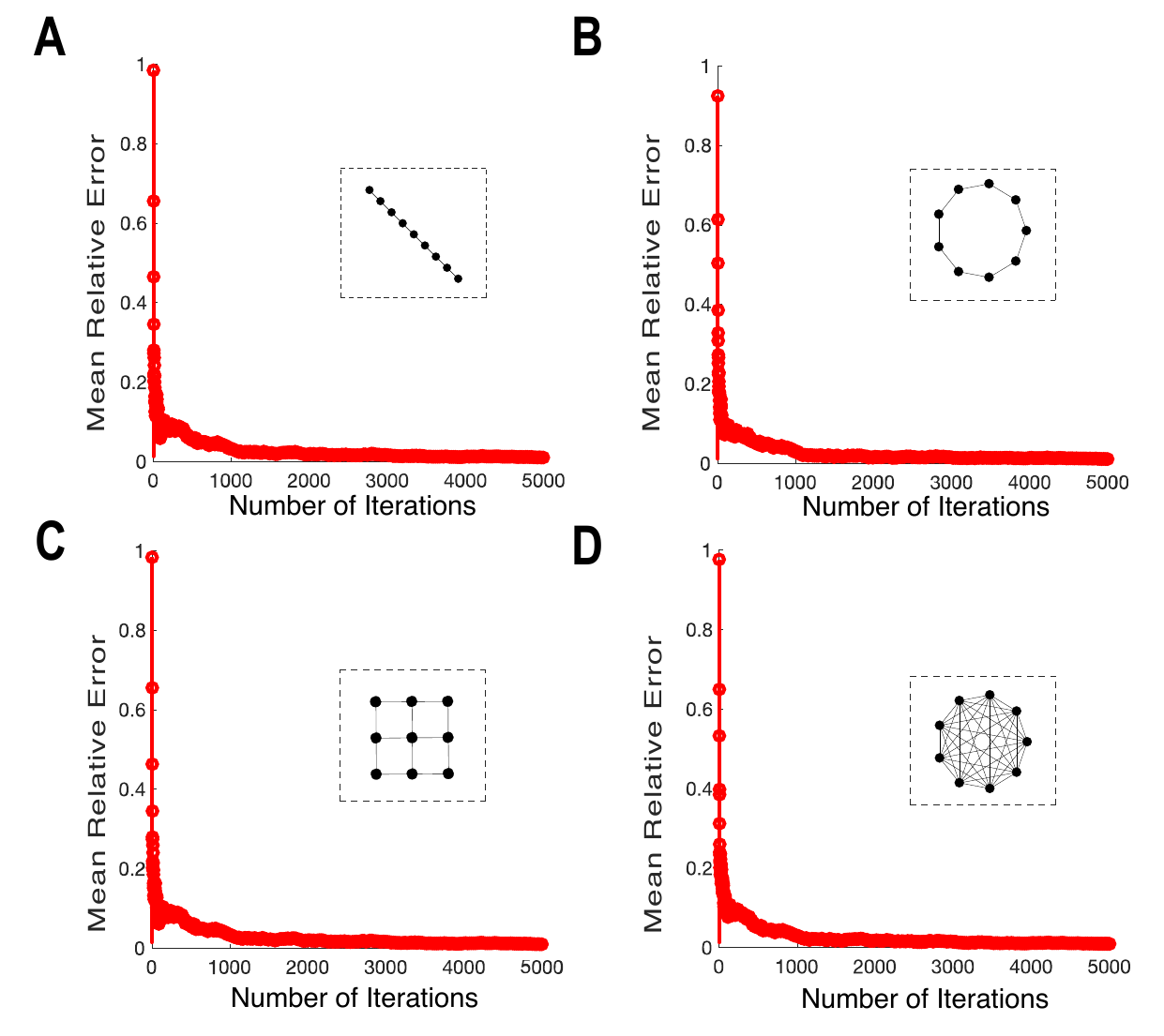}
	\caption{ Recurrent neural network achieves similar inference results as the mean-field method. The mean relative errors rapidly converge in a few iterations.} All MRFs have nine nodes with different topologies as chain (A), single loop (B), grid (C) and fully connected graph (D). $\lambda_1=\lambda_2=0.1$.
	\label{fig:mrf_exp1}
\end{figure}

In order to test the accuracy of the proposed method, we generated several MRFs with different graph topologies (chain, single loop, grid and fully connected graph, see Fig.~\ref{fig:mrf_exp1}), and perform inference of these MRFs with spiking recurrent neural network and mean-field method respectively.

For a MRF with M nodes, we calculated marginal probabilities for all these M nodes with mean-field method and the corresponding recurrent neural network respectively. The mean relative error $\delta$ is defined as follows:
\begin{equation}
\delta = \frac{1}{M}\sum_{i=1}^M\frac{|P^{MF}(x_i=1)-P^{RNN}(x_i=1)|}{P^{MF}(x_i=1)},
\label{eq15}
\end{equation}
where $P^{MF}(x_i=1)$ represents the marginal probabilities computed with mean-field method, and $P^{RNN}(x_i=1)$ represents the result obtained by the corresponding recurrent neural network.

Fig.~\ref{fig:mrf_exp1} illustrates how the relative errors rapidly convergence with a few iterations. For each MRF, the potential functions $J_{ij}$ and $h_i$ are drawn from two uniform distributions on $[0, \lambda_1]$ and $[-\lambda_2, \lambda_2]$ respectively. One can find that even for MRFs with different topologies, the error decreases in a fast way with only a few iterations. These results imply that the simulation of spiking recurrent neural networks can get comparable results as an analytical mean-field method. 

\begin{figure}[t!htp]
	\centering
	\includegraphics[width=0.9\columnwidth]{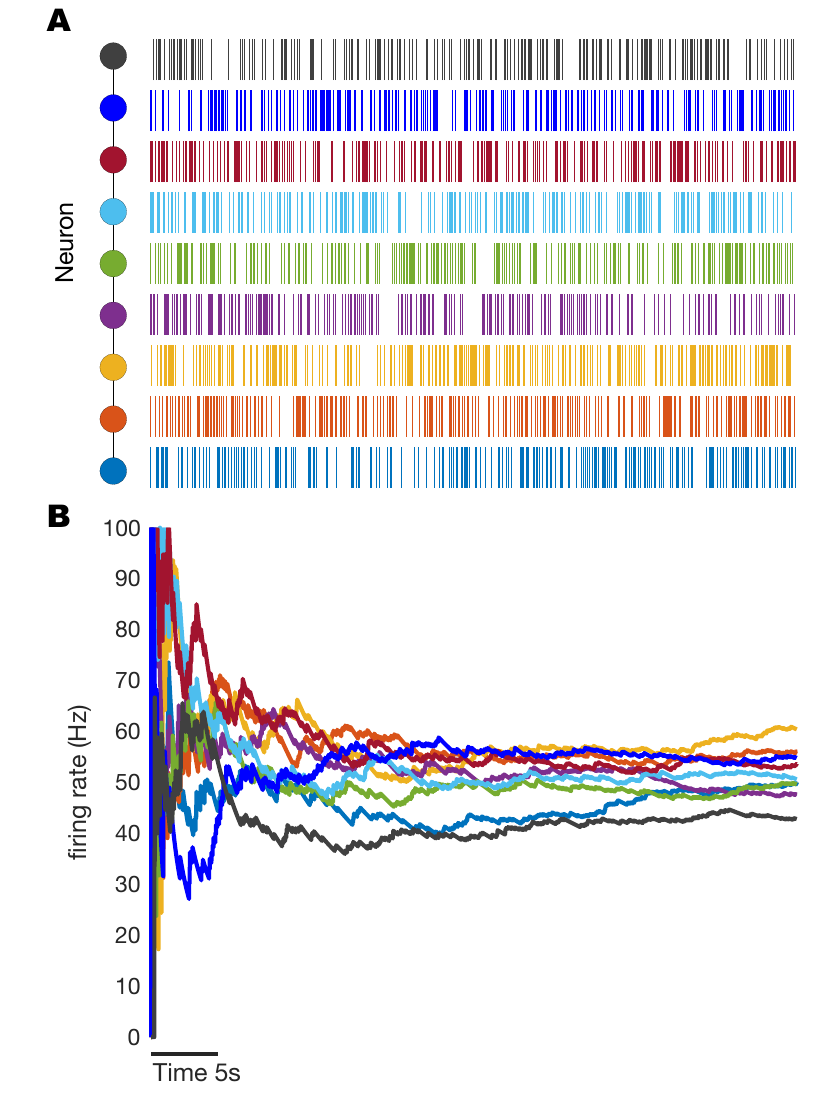}
	\caption{Inference of a chain-structured MRF with spiking recurrent neural network. (A) Spiking activity of the neurons in a recurrent neural network. (B) Time course of firing rates of 9 neurons shown in (A).}\label{fig:mrf_exp1_spk}
\end{figure}

To illustrate the inference mechanism of the spiking recurrent neural network, Fig.~\ref{fig:mrf_exp1_spk}A shows the spiking activity of all 9 neurons in the recurrent neural network when performing inference of a 9-node MRF with chain structure (Fig.~\ref{fig:mrf_exp1}A). Here the mapping between actual firing rate $\hat{r}_i(t)$ and the "firing rate" $r_i(t)$ is $\hat{r}_i(t)=50 r_i(t)+50$. Thus the maximum firing rate of each neuron is $100$ Hz. Fig.~\ref{fig:mrf_exp1_spk}B shows the time course of the firing rate of each neuron. One can see that the firing rate of each neuron converges to a fixed value and then fluctuates around it. 

\subsection{Testing on the Robustness of Our Method}
\begin{figure*}[t!hb]
	\centering
	\includegraphics[width=1.9\columnwidth]{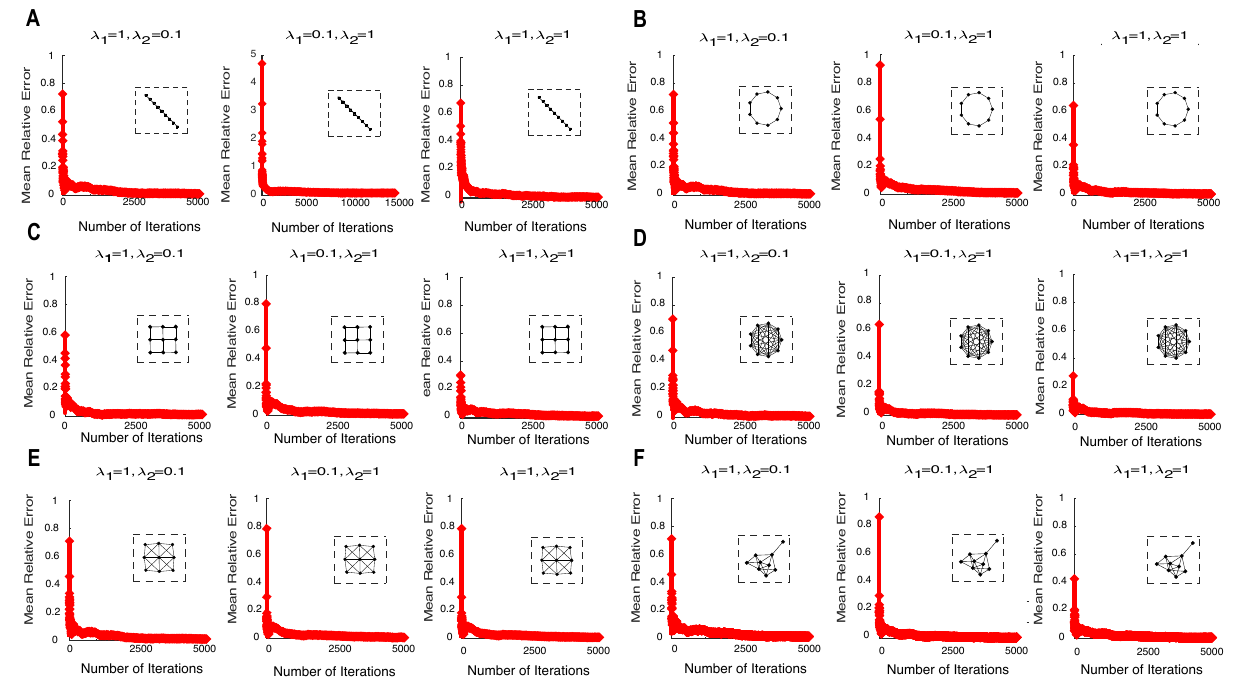}
	\caption{Illustration of the robustness of our method with different settings of parameters on different topologies as chain (A), single loop (B), grid (C), fully connected graph (D), 8-connected grid topology (E) and random connected topology with connection probability 0.5 for each edge (F).
	}\label{fig:mrf_exp1_2}
\end{figure*}

The experimental results above indicate that the inference model of recurrent neural networks can get accurate results as mean-field inference for a given set of parameters of $\lambda_1$ and $\lambda_2$ as 0.1.
Here we make a concrete analysis of the robustness of our model with different parameters. Fig.~\ref{fig:mrf_exp1_2} shows the results where $\lambda_1$ and $\lambda_2$ are set to different combinations of $1$ and $0.1$, except the setting that $\lambda_1=\lambda_2=0.1$ as shown in Fig.~\ref{fig:mrf_exp1}. We can see that, in all cases, the errors converge to almost zero in a fast manner. These results indicate that, different from the previous works \cite{ott2007neurodynamics,yu2017neural} that only apply to MRFs with special potential function ($J_{ij}<<1$ and $h_i<<1$), our method is robust to different parameters and could implement inference for arbitary MRFs.

\begin{figure}[t!hb]
	\centering
	\includegraphics[width=0.8\columnwidth]{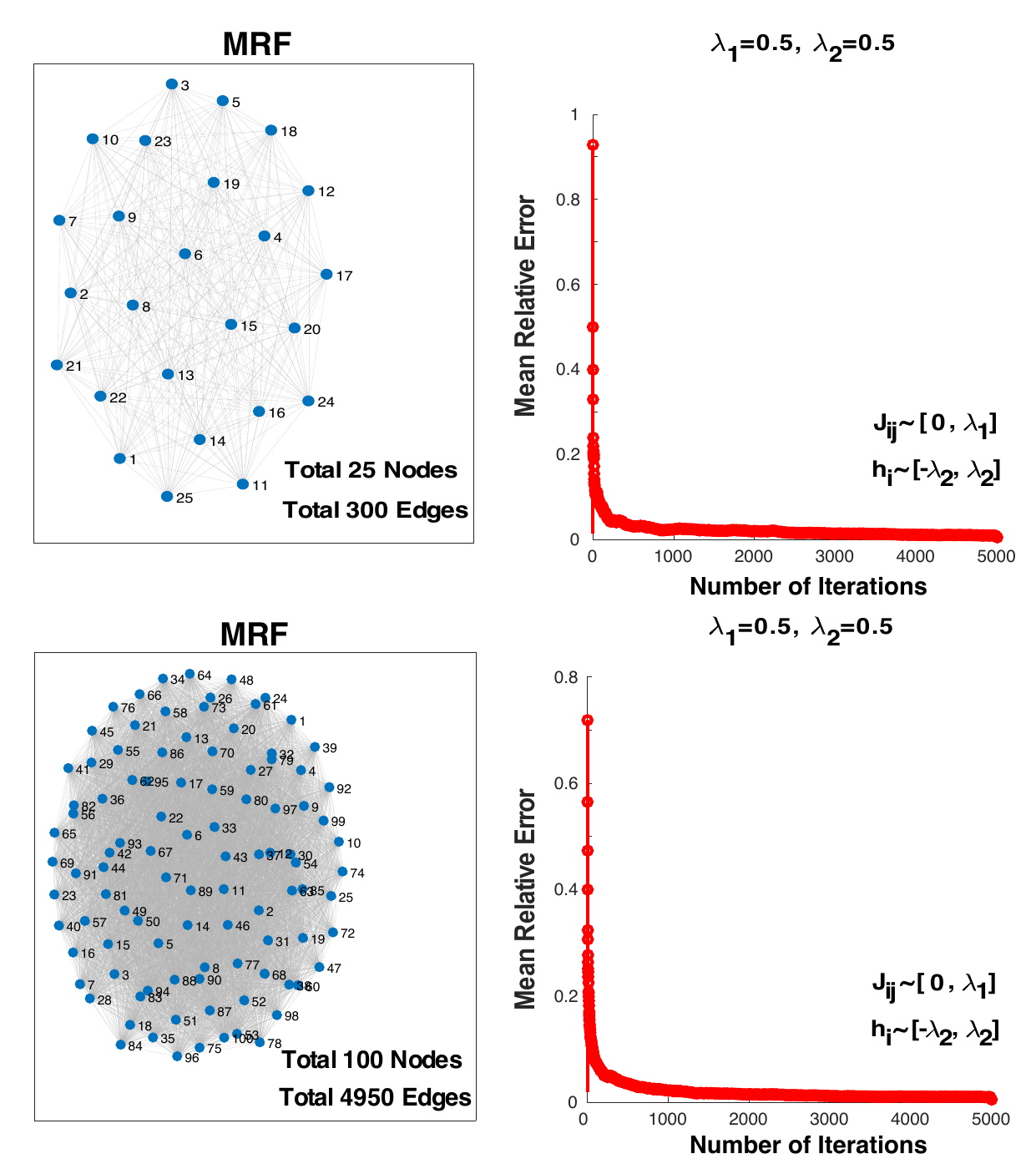}
	\caption{Inference performance of recurrent neural networks on large-scale MRFs.}  \label{fig:mrf_exp2}
\end{figure}

Then we investigate whether our framework can be scaled up to large-scale MRFs with more nodes. Two examples are included here: a MRF with 25 nodes and 300 edges and a MRF with 100 nodes and 4950 edges. As shown in Fig.~\ref{fig:mrf_exp2}, the same conclusion is obtained that the spiking recurrent neural networks can get comparable results as the mean-field method.

\subsection{Binary Images Denoising by Recurrent Neural Networks}

Here we investigate whether our spiking neural network can be scaled up to solve more realistic tasks. We consider the task of image denoising, that is, correcting an image that has been corrupted. In the field of image processing, the researchers often model image denoising problem by  MRFs with grid-like structures (shown in Fig.~\ref{fig:01})  and then convert the denoising problem  to MAP estimation or marginal inference problem. Based on this, we can also tackle this problem with recurrent neural networks by computing the marginal probabilities of each pixel and then infer whether this pixel is white or black in a binary setting.

The image denoising experiments are performed on the NIST Special Database 19 (SD 19), which contains NIST's entire corpus of training materials for handwritten document and character recognition. This dataset includes samples from 3600 writers, consisting of 10 digits $0-9$, 26 lower letters a-z and 26 upper letters A-Z. Therefore we have totally 62 categories. During the experiment, 100 images of each class are randomly selected as dataset. All images used here are $128 \times 128$ pixels. In this experiment, each image is modeled by a square lattice pairwise MRF (shown in Fig.~\ref{fig:01}), where the hidden variables $\{x\}=\{x_1,x_2,...,x_n\}$ represent the denoise image and observed variables $\{y\}=\{y_1,y_2,...,y_n\}$ represent the observed noise image. As observed pixel value is usually the same as the true pixel value, so the unary potential $h(x_i)$ is set to $0.1$ if the variable $x_i$ is the same as the observation variable $y_i$ and $-0.1$ otherwise ($h_i=0.1$). Besides, as nearby pixel values are usually the same in an image, the pairwise potential function $J_{ij}(x_i, x_j)$ is set to 0.8 if $x_i=x_j$ and $-0.8$ otherwise ($J_{ij}=0.8$). 
All the other settings in this experiment are the same as in experiment 4.1 above.

\begin{figure*}[!htbp]
	\centering
	\includegraphics[width=1.8\columnwidth]{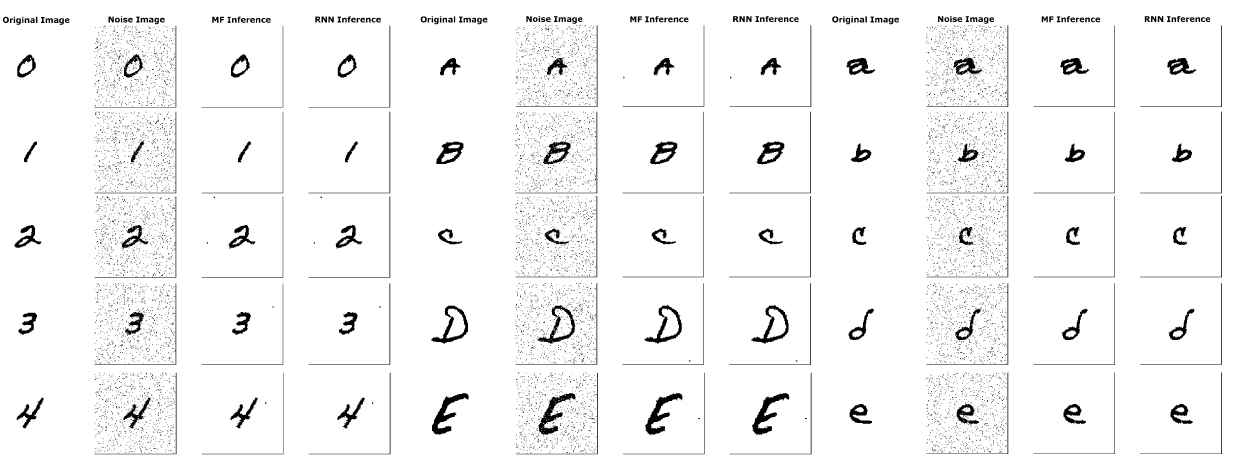}
	\caption{Some examples of image denoising on NIST SD 19 dataset with mean-field inference and recurrent neural networks.}
	\label{fig:nist_denoise}
\end{figure*} 

Fig.~\ref{fig:nist_denoise} shows some examples of image denoising with mean-field inference and the corresponding recurrent neural network. Here the noise images are generated by randomly flipping the pixel value with a probability of $5\%$. 

\begin{figure}[t!htb]
	\centering
    \includegraphics[width=\columnwidth]{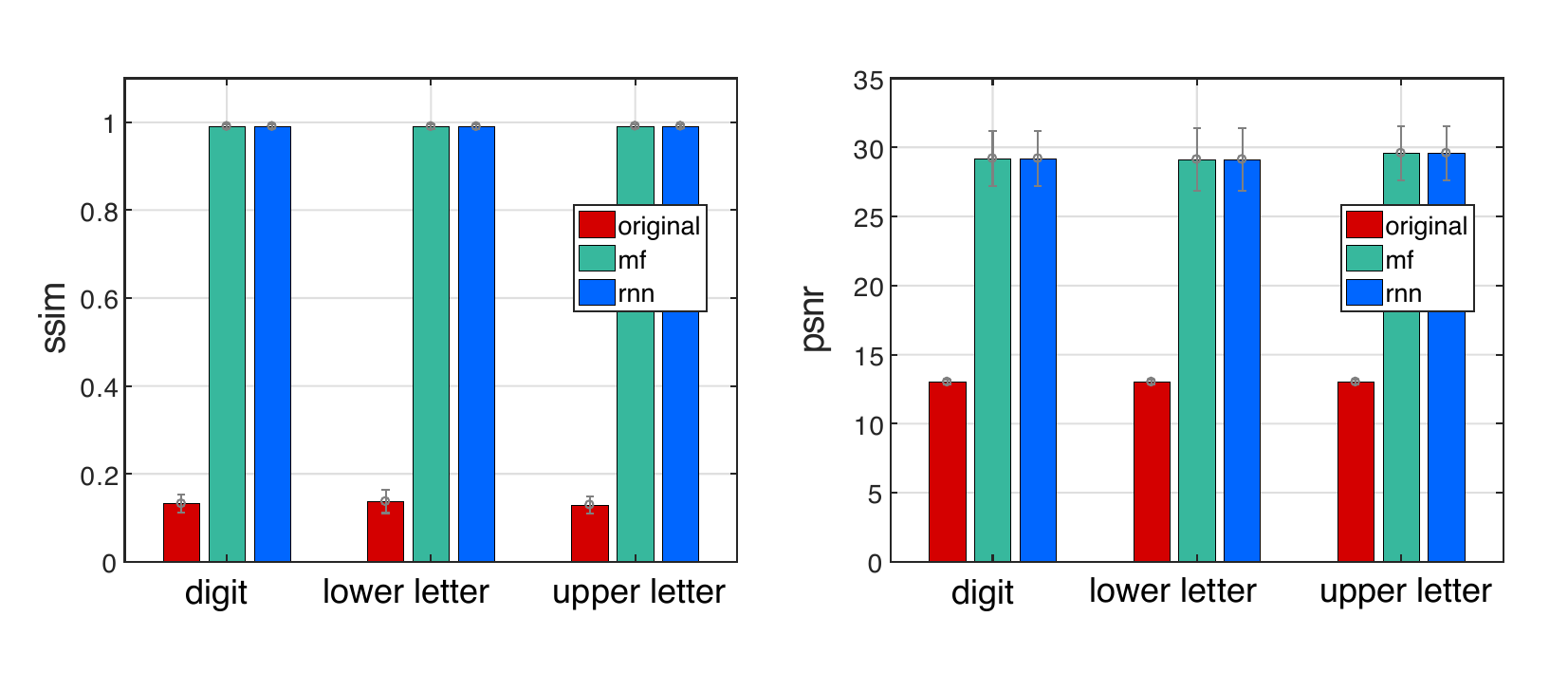}
	\caption{Comparison of denoising result on NIST SD 19 Database with different methods. There are 62 categories from the dataset ($10$ classes for the digit, $26$ kinds of upper letters and $26$ lower letters). Average results over 5 independent trials are shown.}\label{fig:ssim_psnr}
\end{figure} 

\begin{figure}[b!htb]
	\centering
	\includegraphics[width=\columnwidth]{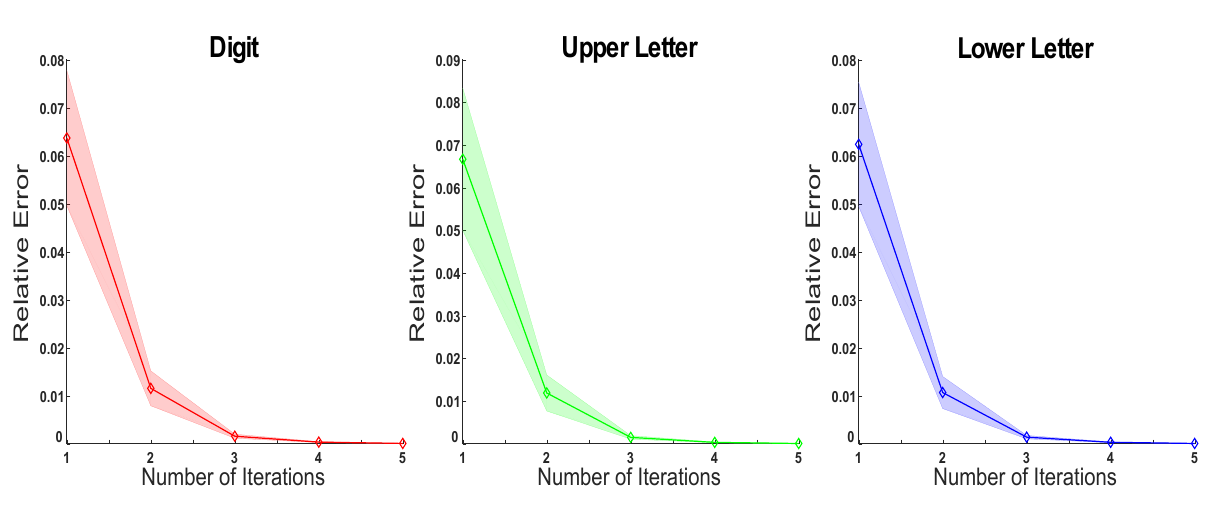}
	\caption{Recurrent neural networks achieve similar results as those of mean-field inference.} The relative error decays to 0 rapidly. The red, green and blue curves represent the results for digit, upper letter and lower letter respectively. All the results are averaged over 5 independent trials, shaded area indicates standard deviation (STD).
	\label{fig:error_res}
\end{figure}

We also quantitatively analyze these results by computing the structural similarity index (SSIM) and the peak signal-to-noise ratio (PSNR). As shown in Fig.~\ref{fig:ssim_psnr}, the SSIM of the original image, denoised image by mean-field inference, and denoised image by recurrent neural networks are $13.01 \pm 0.14$, $29.19 \pm 1.97$ and $29.19 \pm 1.97$, respectively. The PSNR of the original image, denoised image by mean-field inference, and denoised image by recurrent neural networks are $0.1322 \pm 0.0207$, $0.9905 \pm 0.0047$ and $0.9905 \pm 0.0047$ respectively. All these results demonstrate that recurrent neural networks can get the same denoising results as mean-field inference. Fig.~\ref{fig:error_res} illustrates how the mean relative error between recurrent neural networks and mean-field inference varies over time. We can find the error converges to 0 with a few iterations.

\subsection{Comparison among Different Neural Network Based Image Denoising Methods}

\begin{figure}[t!htb]
	\centering
	\includegraphics[width=0.8\columnwidth]{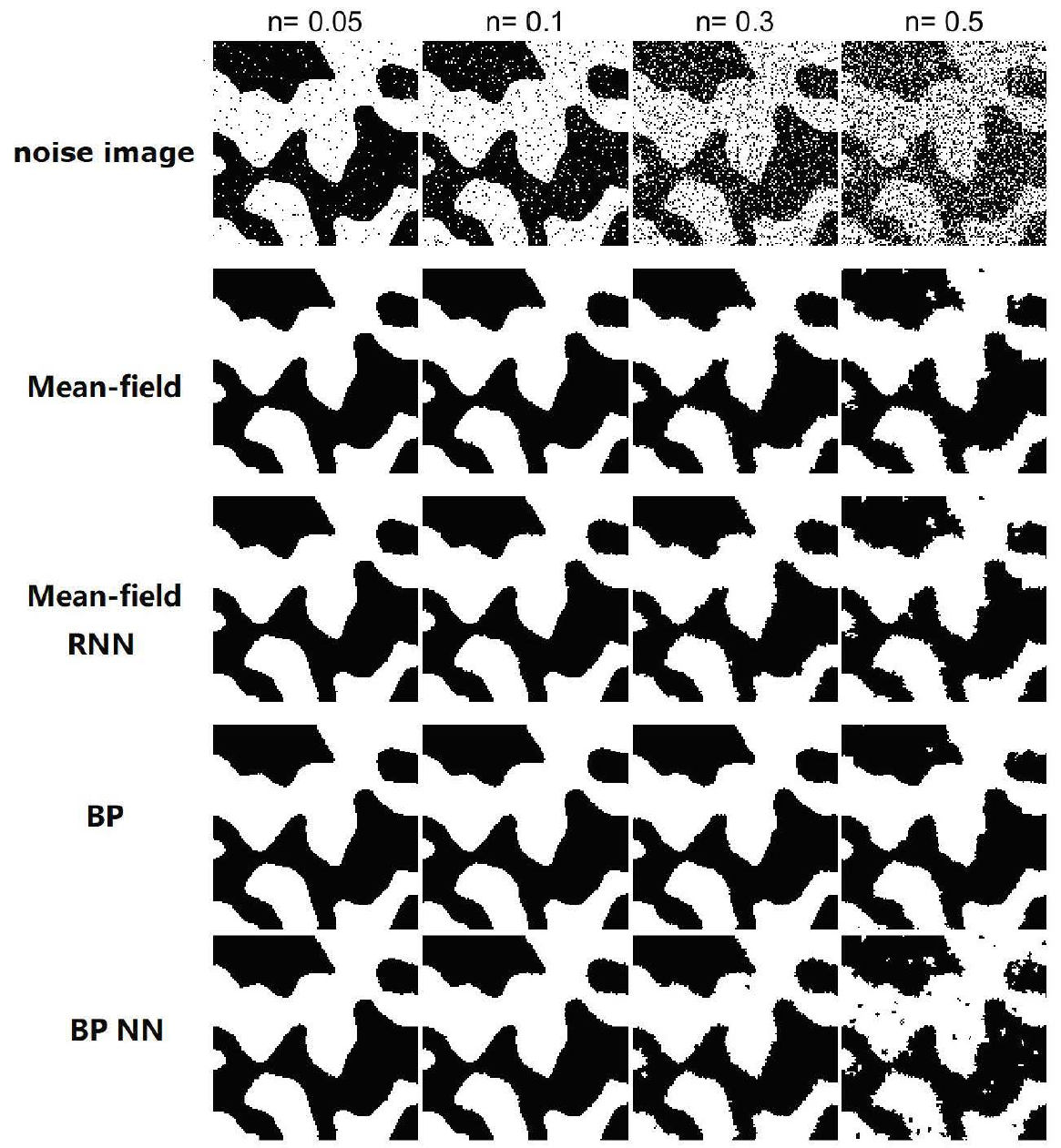}
	\caption{Image denoising with mean-field inference, recurrent neural networks, BP algorithm and BP-based neural networks. Here $n$ denotes different noise levels.}
	\label{fig:random_denoise_samples}
\end{figure}

In section 3.4, we have proved that the previous approaches based on BP and reparameterization (BP-based neural networks) can be unified in our framework. In order to test this, we compare our method with the BP algorithm and the BP-based neural network model for the task of image denoising. In order to increase the difficulty of inference, here we created a dataset of 100 images with $128 \times 128$ pixels by making randomly noisy images and then smooth them to get true output values. Fig.~\ref{fig:random_denoise_samples} shows one example of the randomly generated binary images. One can find that there exists more separated space in these images compared with the images in NIST SD 19. Thus it's more difficult to be denoised. 

\begin{figure}[t!htb]
	\centering
	\includegraphics[width=\columnwidth]{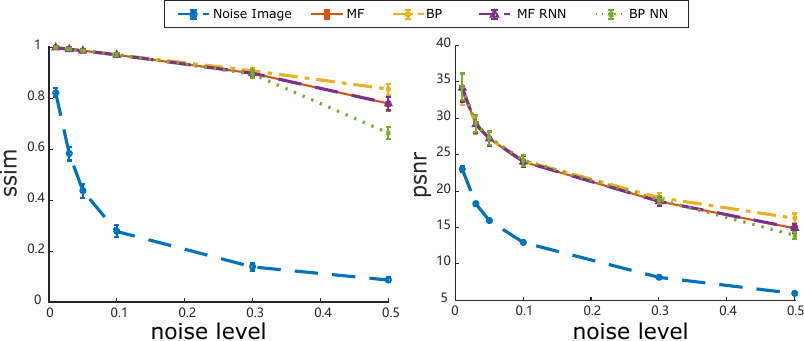}
	\caption{Comparison of the performance of difference denoising methods with respect to noise level. }\label{fig:random_denoise_res}
\end{figure}  

To compare the performance of these algorithms, we add the different levels of \emph{salt and pepper} noise on the binary images, and characterize the quality of the denoised images with the criterion of the SSIM and PSNR. Fig.~\ref{fig:random_denoise_samples} illustrates one example of image denoising with mean-field inference, recurrent neural networks, BP algorithm \cite{koller2009probabilistic}, and BP-based neural networks \cite{yu2017neural}.
Fig.~\ref{fig:random_denoise_res} compares the SSIM and PSNR of different methods and noise levels. One can see that the performance of recurrent neural networks (purple curve) is the same as mean-field inference (red curve). Besides, one can also find that the performance of BP-based neural networks (green curve) is nearly the same as that of recurrent neural networks (purple curve) and mean-field inference (red curve), which demonstrates that the previous work equals mean-field inference and can be unified in our framework. Note that there exists a gap between BP-based  neural  networks  (green  curve)  and recurrent neural networks (purple curve) when the noise level is larger than 0.3, which comes from the approximation between Eq. \eqref{BP2} and Eq. \eqref{BP3}.

\section{Conclusion}
In this paper, we prove that there exists a precise equivalence between the dynamics of recurrent neural network and mean-field inference of binary Markov random fields. We show that if the synaptic weights and input current encode the potential function of MRFs, the firing rates of neuron in recurrent neural networks encode the difference between the probabilities for two states. The time course of neuronal firing rate can implement marginal inference. Theoretical analysis and experiments on MRFs with different topologies show that our neural network can get the same performance as the mean-field method. Besides, we also apply our proposed spiking framework to practical computer vision problem, i.e., binary images denoising. 

Differ from previous works based on BP algorithm and reparameterization, where  the potential functions of MRF should meet some strict conditions, we design a spiking network that can implement mean-field inference for arbitrary MRFs. What's more, we have demonstrated that our work unifies previous works. 

The previous work of neural implementation of Bayesian inference \cite{rao2004bayesian,deneve2008bayesian,yu2018unification} with recurrent neural networks focused on inference of hidden Markov models. There also exist some studies \cite{rao2005hierarchical,guo2017hierarchical} that extended the networks to a multilayer structure to perform hierarchical Bayesian inference. Different from these works, we are focusing on how spiking neural networks are able to implement probabilistic inference of MRF. In future work, we will try to extend our proposed framework to tackle more advanced realistic problems, like recognition and stereo matching. 

\section*{Acknowledge}
This work is supported in part by the National Natural Science Foundation of China under grants 61806011 and 61825101, in part by National Postdoctoral Program for Innovative Talents under grant BX20180005, in part by China Postdoctoral Science Foundation under grant 2018M630036, in part by the Zhejiang Lab under grants 2019KC0AB03 and 2019KC0AD02,
in part by the Royal Society Newton Advanced Fellowship under grant NAF-R1-191082.

\bibliographystyle{plain}
\bibliography{pibmrf}

\end{document}